\documentclass[aps,prb,reprint,twocolumn,letterpaper,floatfix,preprintnumbers,showpacs]{revtex4-1}

\usepackage{graphicx}
\usepackage{dcolumn}
\usepackage{bm}
\usepackage{color}
\usepackage{epsfig}
\usepackage{latexsym}
\usepackage{amsmath}
\usepackage{xspace}
\usepackage[normalem]{ulem}

\newcommand{\fullsoc}{$G^{\mathrm{SOC}}W^{\mathrm{SOC}}$\xspace}
\newcommand{\bise}{Bi$_2$Se$_3$\xspace}

\newcommand{\bisb}{Bi$_{1-x}$Sb$_x$\xspace}
\newcommand{\kv}{\textbf{k}\xspace}
\newcommand{\abinitio}{\textit{ab initio}\xspace}
\newcommand{\eg}{e.g.,\xspace}
\newcommand{\ie}{i.e.,\xspace}

\begin{document}

\title{Electronic phase transitions of bismuth under strain from relativistic self-consistent \textit{GW} calculations}

\author{Irene Aguilera, Christoph Friedrich, Stefan Bl\"{u}gel}

\address{Peter Gr{\"{u}}nberg Institute and Institute for Advanced Simulation, Forschungszentrum J{\"{u}}lich and JARA, 52425 J{\"{u}}lich, Germany}


\date{\today}

\begin{abstract}
We present quasiparticle self-consistent $GW$ (QS$GW$) calculations of
semimetallic bulk Bi. 
We go beyond the conventional QS$GW$ method 
by including the spin-orbit coupling throughout the self-consistency cycle. 
This approach improves the description 
of the electron and the hole pockets considerably with respect to
standard density functional theory (DFT), leading to excellent agreement with experiment.
We employ this relativistic QS$GW$ approach to conduct a study of the
semimetal-to-semiconductor and the trivial-to-topological 
transitions that Bi experiences under strain. 
DFT predicts that an unphysically large strain is needed for such transitions. We 
show, by means of the relativistic QS$GW$ description of the electronic structure, that an 
in-plane tensile strain of only 0.3\% and a compressive strain of 0.4\% are sufficient to cause the 
semimetal-to-semiconductor and the trivial-to-topological phase transitions, respectively. Thus,
the required strain moves into a regime that is likely to be realizable in experiment, which 
opens up the possibility to explore bulklike topological behavior of pure Bi.
\end{abstract}

\maketitle

Bismuth exhibits a series of peculiarities that has made it the subject of experimental and theoretical interest 
for decades. 
The very low density of carriers with 
high carrier mobility and a very long mean free path make 
bismuth very interesting for electronic transport studies. 
With the recent introduction of the classification of solids according to topology classes of the electronic
structure,\cite{moore2010} Bi has again moved into the limelight of current discussions. With a lattice structure that bears similarities to
graphene and a very large spin-orbit interaction, it has all the potential to be a topological semimetal or semiconductor.
From the theoretical point of view, the 
large spin-orbit coupling (SOC), the semimetallic character, and the very small local direct band gap 
at the L point of the Brillouin zone make \abinitio calculations 
of the electronic structure of Bi very challenging. 

As for its topological classification, some important aspects 
are not yet fully understood. It is widely accepted that bulk Bi is 
topologically trivial in contrast to Sb. 
This is in agreement with angle-resolved
photoemission spectroscopy (ARPES) measurements of Bi\cite{ast2003,koroteev2004,hofmann2006} and \bisb 
alloys,\cite{hsieh2008,roushan2009,hsieh2009} as well as transport studies, 
in which a semimetal-to-semiconductor (SMSC)\cite{tichovolsky1969} and a 
topological transition\cite{redko1985} are observed as the Sb concentration is varied from the topologically trivial Bi to the 
topological semimetal Sb, resulting in three-dimensional (3D) topological insulators for some of the
alloys. 
However, controversies concerning the topological properties of bismuth 
persist. For example, it has recently been claimed\cite{ohtsubo2013} on the basis of ARPES measurements 
that bulk bismuth is topologically nontrivial, a claim that 
is in conflict with 
most previous findings including \abinitio calculations based on density functional theory (DFT). 
The authors of Ref.~\onlinecite{ohtsubo2013} attributed these discrepancies
to the systematic underestimation of the band gaps obtained 
in the local-density (LDA) or generalized gradient (GGA) approximations of DFT.
Given the very small experimental value 
(11-15~meV\cite{maltz1970,vecchi1974,isaacson1969,brown1963,smith1964})
for the direct band gap at the L point ($E_\mathrm{g}$), whose ``sign'' 
(order of the states) controls the topological nature of Bi, 
it is very conceivable that DFT might incorrectly predict the sign of the
band gap, and not just its magnitude. Our parity analysis within LDA is 
in agreement with results in the literature (see,
\eg Refs.~\onlinecite{golin1968,timrov2012,teo2008}) and 
confirms the trivial character of Bi. But surprisingly, LDA presents an unusual 
\textit{overestimation} of the band gap at L (86~meV; see Table~\ref{tablegaps}), which 
is an effect usually characteristic of inverted gaps, as with those in topological 
insulators.\cite{yazyev2012,aguilera2013}  
The gap at L is indeed inverted, as can be seen by varying the SOC strength 
from zero to 100\%: the two states at L exchange order for a certain SOC 
strength. But the gap at T is inverted as well. There is, hence,
an even number of inverted gaps at the time-reversal invariant momenta,
giving rise to a topologically trivial value of the $Z_2$ invariant.\cite{kane2005-1,fu2007}
The GGA approximation hardly improves the LDA results (Table~\ref{tablegaps}). 
The poor description of the band gaps casts doubt on the LDA and GGA studies of
the trivial-to-topological (TT)\cite{hirahara2012,liu2011-2} or the SMSC\cite{gutierrez2006,hirahara2012} 
transitions of Bi, as well as the TT transition in \bisb alloys, as these 
studies critically depend on the sign and the value of the band gap at L. 

The cause of the band-gap problem of DFT is well known. Strictly, the
single-particle energies cannot be interpreted as quasiparticle energies of
the interacting electron system. The $GW$ approximation\cite{hedin1965} for the electronic self-energy, applied as 
a one-shot correction to DFT or, more rarely, in the quasiparticle
self-consistent $GW$ (QS$GW$) approach,\cite{faleev2004} 
remedies the aforementioned problem and is often used to correct
the band gaps obtained from LDA or GGA. In most cases 
it increases the band gap leading to a much better agreement with
experiments.\cite{schilfgaarde2006} However, we show here ({\it cf.}\ Table~\ref{tablegaps}) that in the 
case of bismuth, the $GW$ quasiparticle correction is
negative, i.e., it correctly \textit{reduces} the band gap 
to 32~meV in one-shot $GW$ (which corresponds to the 
first iteration of QS$GW$) and to 13~meV within the self-consistent QS$GW$ scheme. 
The latter value lies in the range of experimental values. 
For these calculations, we
have devised and implemented a method that combines the SOC with the many-body
renormalization in a consistent way (see below). The sign of the
band gap is not changed by the quasiparticle correction, \ie bismuth remains
trivial, in accordance with the vast majority of experimental and theoretical
studies. However, the phase boundary at which 
bismuth becomes topologically nontrivial is considerably closer in
QS$GW$ than in LDA.
We will show that only a relatively modest in-plane 
tensile strain of 0.3\% and compressive strain of 0.4\% 
are sufficient to cause the SMSC and TT phase transition, 
respectively, which opens up the possibility to explore the 3D 
topological properties of \textit{pure} Bi.

\begin{table}
\caption{\label{tablegaps} Values of the direct band gap at L ($E_\mathrm{g}$) and the indirect band gap ($E_0$) [see Fig.~\ref{figbands}(b)] for bulk bismuth (in meV) calculated within LDA, GGA, $GW$, and QS$GW$, compared with experimental results.}
\begin{tabular}{rccccc}
\hline
\hline
 & LDA & GGA & $GW$ & QS$GW$ & Expt. \\
\hline
direct gap at L ($E_\mathrm{g}$)  &   \phantom{$-$1}86 &  \phantom{$-$}74 &  \phantom{$-$}32 & \phantom{$-$}13 &  \phantom{$-$}11 -- \phantom{$-$}15\footnotetext[1]{11.0~[\onlinecite{maltz1970}], 13.6~[\onlinecite{vecchi1974}], 15.0~[\onlinecite{isaacson1969,brown1963}], 15.3~[\onlinecite{smith1964}]}\footnotemark[1] \\
indirect gap T-L ($E_0$) & $-$105 & $-$72 & $-$66 & $-$33 & $-$32 -- $-$39\footnotetext[2]{-32.0~[\onlinecite{edelman1975}], -36.0~[\onlinecite{isaacson1969}], -38.0~[\onlinecite{dinger1973}], -38.2~[\onlinecite{isaacson1969}],-38.5~[\onlinecite{smith1964}]}\footnotemark[2] \\
\hline
\hline
\end{tabular} \\
\end{table}

\begin{figure}
\begin{center}
\includegraphics[angle=90,width=8.2cm]{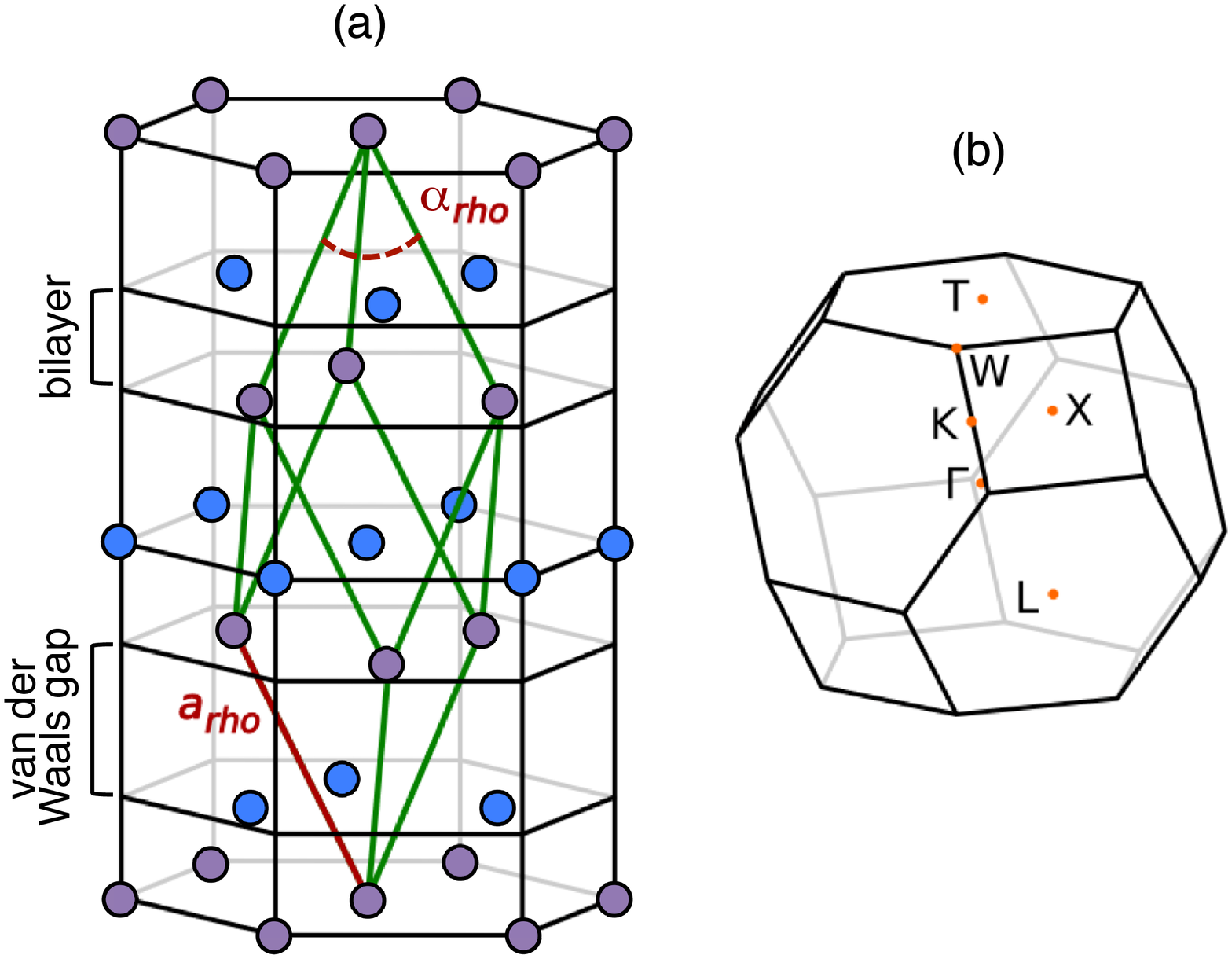}
\caption{\label{struct} (Color online) (a) Rhombohedral (green) unit cell and hexagonal crystal structure of bulk Bi. 
The blue and purple atoms distinguish the atoms in the two layers of each bilayer. (b) Bulk Brillouin zone of Bi expressed in the rhombohedral unit cell.}
\end{center}
\end{figure}

\begin{figure}
\begin{center}
\includegraphics[angle=0,width=8.2cm]{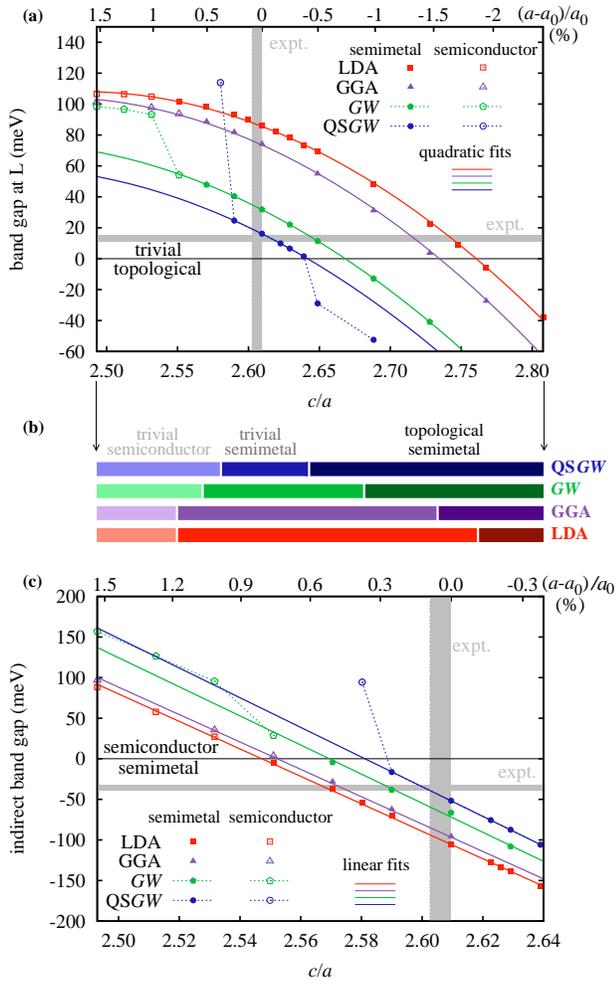}
\caption{\label{figstrain} (Color online) (a) LDA, GGA, $GW$, and QS$GW$
direct band gap $E_\mathrm{g}$ of bismuth at L for different strains (varying the
ratio $c/a$ and keeping the volume constant). 
(b) Three electronic phases are found for Bi as a function of the ratio $c/a$ 
within all the approaches. 
(c) LDA, GGA, $GW$, and QS$GW$ indirect band gap $E_0$ of bismuth for different values of $c/a$. 
On the upper $x$ axis of (a) and (c), we show the percentage of in-plane strain. 
The range of experimental values of $c/a$,\cite{schiferl1969} 
the experimental gap at L,\cite{maltz1970,vecchi1974,isaacson1969,brown1963,smith1964} 
as well as the experimental indirect gap\cite{edelman1975,dinger1973,isaacson1969,smith1964} 
are represented as gray shaded areas in (a) and (c). Small differences to the QS$GW$ values 
in Table~\ref{tablegaps} are due to the use of a reduced \kv-point set; see Appendix.}
\end{center}
\end{figure}

Bulk bismuth crystallizes in the A7 rhombohedral structure with R$\bar 3$m space group and two atoms per unit cell.
The cell and the atomic positions are determined by the internal parameter $u$, the rhombohedral lattice 
parameter $a_\mathrm{rho}$, and the rhombohedral angle $\alpha_\mathrm{rho}$ [see Fig.~\ref{struct}(a)]. The layered structure consists of Bi bilayers
separated by a van der Waals gap.
For the undistorted reference lattice, we employ
experimental parameters from Ref.~\onlinecite{schiferl1969}: $a_\mathrm{rho}$=4.7458~\AA{}, $\alpha_\mathrm{rho}$=57.230$^{\circ}$,
$u$=0.23389.
Equivalently, the structure can be described as hexagonal with in-plane and
out-of-plane lattice constants $a_0$=4.5460~\AA{} and $c_0$=11.862~\AA{} ($c_0/a_0$=2.6093). 
These lattice parameters were measured at 298~K. The gray vertical areas  
in Fig.~\ref{figstrain} show the range of experimental values of $c/a$ for temperatures 
between 4 and 298~K.\cite{schiferl1969} 
All of our calculations are carried out with the DFT code {\sc fleur}{}\cite{fleur} and the
$GW$ code {\sc spex}{},\cite{friedrich2010-spex} realizations in the all-electron full-potential linearized
augmented-plane-wave (FLAPW) formalism.
The convergence parameters of the DFT, $GW$, and QS$GW$ calculations are given in 
the Appendix. To include relativistic effects in the DFT
calculations, we employ the scalar-relativistic\cite{koelling1977} 
approximation inside the muffin-tin spheres, and the SOC is included in
a self-consistent manner.\cite{li1990}

Most of the one-shot $GW$ calculations that include SOC
and all of the QS$GW$ ones published so far, employ the SOC as an \textit{a posteriori} correction
after the quasiparticle correction to DFT has been carried out without SOC.
This is, of course, an approximation. For example, many-body renormalization effects of
spin-orbit split bands are then
not taken into account (see Ref.~\onlinecite{aguilera2013-2} for a detailed
discussion). In our relativistic calculations, the SOC is already incorporated in the noninteracting
system that serves as the starting point for the quasiparticle
calculation.\cite{aguilera2013-2,sakuma2011}
The one-shot $GW$ self-energy thus acquires terms that couple the two spin channels, enabling a many-body
renormalization of the SOC itself.
In this work, 
we propose to extend this principle 
to the self-consistent QS$GW$
calculations, so that the self-energy contains spin off-diagonal blocks that it inherits
from the SOC throughout the whole self-consistent process.
(In analogy to the notation \fullsoc of Ref.~\onlinecite{aguilera2013-2}, one could denote
this approach by QS\fullsoc, but for simplicity we will simply write $GW$ and QS$GW$ in the following.)
Our results show that this treatment is crucial for Bi. The
\textit{a posteriori} correction mentioned above, on the other hand,
leads to unphysical results for Bi (see Appendix),
which then appears as a topological insulator with large band gaps
(83 and 259~meV in one-shot $GW$ and QS$GW$, respectively)
instead of a trivial semimetal. 

\begin{figure}
\begin{center}
\includegraphics[angle=0,width=8.2cm]{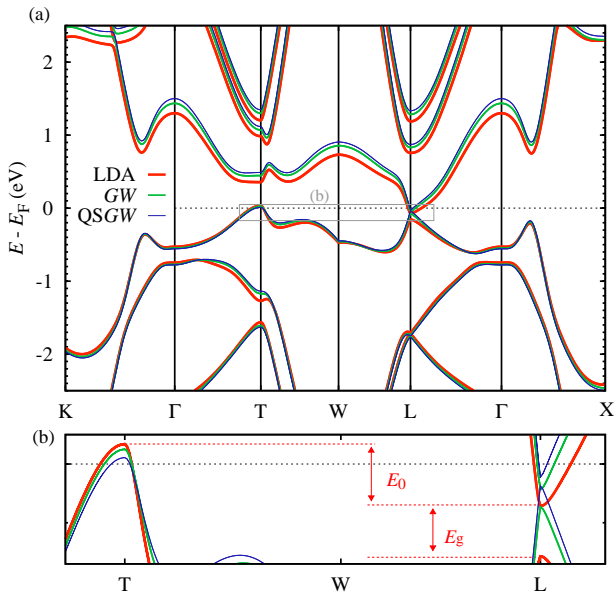}
\caption{\label{figbands} (Color online) LDA, $GW$, and QS$GW$ band structures of bulk Bi. 
The inset in (b) shows the electron and hole pockets together with the $E_g$ and $E_0$ 
gaps corresponding to the LDA bands.} 
\end{center}
\end{figure}

In addition to the value of the band gap at the L point, the properties of Bi are also determined by the
states at the symmetry point T. In particular, the semimetallic character of bismuth is caused by the
overlap between the highest valence band at T and the lowest conduction band at L [see Fig.~\ref{figbands}(b)]. This creates three
very small electron pockets (at the three L points in the Brillouin zone) and one hole pocket at T.
Several attempts to describe these pockets theoretically have been published in recent years.
For a review and comparison of them see, \eg Table II in Ref.~\onlinecite{timrov2012}. The approaches
used include LDA and GGA,\cite{timrov2012,gonze1988,shick1999} tight-binding models,\cite{liu1995} and
empirical pseudopotential calculations,\cite{golin1968} but no $GW$ results
have been published so far.
The results from the models and DFT calculations 
have not been satisfactory: the absolute value of the overlap, or 
in other words the indirect band gap
$E_0$, came out too
large in general compared to experiment.

Our values of $E_0$ 
are shown in Table~\ref{tablegaps}. Both the LDA and GGA overestimate indeed this indirect band gap, in line with previous
publications.
While the one-shot $GW$ approach does correct $E_0$ in the
right direction, it uses LDA as a starting point and, thus, inherits some of
this overestimation. The self-consistent procedure of QS$GW$, on the other
hand, makes the result independent of the starting point and
moves $E_0$ inside the range of experimental values.

Figure~\ref{figbands} shows the band structure of bulk bismuth obtained with
LDA, $GW$, and QS$GW$ along
the path connecting the high-symmetry \kv points represented in Fig.~\ref{struct}(b).
Whereas the overall shape of the band structure remains basically unaffected by the many-body corrections,
there are important changes in the details. In particular, while in most
parts the valence and conduction bands are shifted apart by the many-body corrections as commonly seen in $GW$
calculations, the changes of the dispersion in the vicinity of the Fermi energy
are such that the absolute values of $E_\mathrm{g}$ and $E_0$ get smaller and 
reach values of $13$ and $-33$~meV, respectively. As has
already been discussed before,\cite{sakuma2011,yazyev2012,aguilera2013,aguilera2013-2} 
this is the result of a delicate interplay between the SOC and the many-body
renormalization described by the $GW$ approximation. We conclude 
that the present relativistic QS$GW$ is the method of choice to study 
the band structure of Bi and its topological character
as well as the TT and SMSC transitions that Bi experiences under strain.

A TT transition was indeed recently observed experimentally for strained 
Bi(111) ultrathin films,\cite{hirahara2012} and a
similar transition was found for strained Bi nanowires elongated about 2\%.\cite{condrea2013}
In the case of thin films, a strain can be introduced naturally by a lattice 
mismatch between film and substrate. 
To simulate the strain in our calculations, we apply a volume-conserving distortion of the 
structure by varying the $c/a$ ratio, and keeping the internal parameter $u$ constant. 
A compression in the $c$ direction is accompanied by a
corresponding increase of $a$, and vice versa. This is equivalent to varying the 
rhombohedral angle $\alpha_\mathrm{rho}$, see Fig.~\ref{struct}(a). 
As the $c/a$ ratio increases,
the nearest-neighbor distance reduces and the van
der Waals gap expands. Both effects lead to a
strengthening of the in-plane interactions,
while the interbilayer coupling weakens. 
The altered crystal field reduces the band gap at L until it reaches a zero value
at a critical $c/a$ ratio and becomes negative. This undoes the band inversion 
and causes a sign change
in the calculation of the
$Z_2$ topological invariants,\cite{kane2005-1,fu2007} which, for systems with inversion symmetry, 
derive from the parities of the occupied states at the time-reversal invariant momenta. 
The parity of a wave function remains unchanged upon applying the $GW$
quasiparticle correction. The $Z_2$ topological invariants
can, thus, only change if there is an interchange of valence and
conduction states with respect to LDA. 
A topological-to-trivial transition in the topological insulator \bise as a result of strain has also been discussed in
the literature. It has been argued\cite{young2011,liu2011} that tensile out-of-plane strain reduces 
the spin-orbit strength leading to a trivial phase. We do not find such an effect for bulk
Bi: no substantial increase or decrease in the spin-orbit strength is observed upon application of strain.

Figure~\ref{figstrain}(a) shows the direct band gap as a function of the
$c/a$ ratio with the range of equilibrium experimental values of bulk Bi represented as gray shaded areas.
Positive values of the gap denote a trivial phase, whereas negative values
correspond to a topologically nontrivial case. In addition to the
semimetallic state (solid symbols)
in Fig.~\ref{figstrain}(a), we also observe a transition to a semiconducting
state (open symbols) for tensile in-plane strain. 
This SMSC transition can also be seen in Fig.~\ref{figstrain}(c) showing the variation 
of the indirect band gap, which changes sign and becomes positive at the point of transition. 

At the SMSC transition, the $GW$ and QS$GW$ curves exhibit abnormal
behavior; there is a jump (dashed lines) where we expect the curve to be continuous. An
analogous steplike behavior can be seen in the QS$GW$ curve at the TT
transition. As shown in the Appendix, these anomalies are an artifact of
the \kv-point convergence: The steps
get systematically smaller when one improves the \kv-point sampling, and
convergence is achieved much more readily in the trivial 
semimetallic phase than in the other two phases, so that we can take the 
points between the transitions in Fig.~\ref{figstrain} as converged, and the
fitted curves (solid lines) show the expected behavior in the sampling limit.
We have employed a quadratic function for the direct gap and a linear function
for the indirect gap as a function of $c/a$, which for QS$GW$ give
$E_\mathrm{g}=-1226.28(c/a)^2+5939.05(c/a)-7131.51$ and
$E_0=-1827.57(c/a)+4717.26$ in meV.

\begin{table}
\caption{\label{tabletrans} Values of the lattice parameters $a$ and $c$ and
the ratio $a/c$ for which (a) the TT and (b) the SMSC transitions occur for the different
theoretical approaches. The values in brackets indicate the compressive 
($-$) or tensile (+) strain in percentage.
The difference in the total energies ($\Delta E_\mathrm{TOT}$) between the structure
for which the transition occurs and the unperturbed reference lattice has
been calculated within LDA. Note that we used
$a_0$=4.5460~\AA{}, $c_0$=11.862~\AA{}, and 
$c_0/a_0$=2.6093\cite{schiferl1969} for the undistorted lattice.} 
\begin{tabular}{lcccc}
\hline
\hline
 & $a_\mathrm{t}$ & $c_\mathrm{t}$ & $c_\mathrm{t}/a_\mathrm{t}$ & $\Delta E_\mathrm{TOT}$  \\
 & (\AA) & (\AA) &  & (meV)  \\
\hline 
(a) \textbf{TT:} & & & & \\
\hline
LDA    & 4.461($-$1.9\%) & 12.315 & 2.761(+5.8\%) & 15.8  \\
GGA    & 4.475($-$1.6\%) & 12.238 & 2.735(+4.8\%) & 11.7  \\
$GW$   & 4.513($-$0.7\%) & 12.036 & 2.667(+2.2\%) & 4.1  \\
QS$GW$ & 4.527($-$0.4\%) & 11.960 & 2.642(+1.2\%) & 1.9  \\
\hline
(b) \textbf{SMSC:} & & & & \\
\hline
LDA    & 4.579(+0.7\%) & 11.690 & 2.553($-$2.2\%) & 0.5  \\
GGA    & 4.580(+0.7\%) & 11.685 & 2.551($-$2.2\%) & 0.6  \\
$GW$   & 4.569(+0.5\%) & 11.744 & 2.570($-$1.5\%) & 0.2  \\
QS$GW$ & 4.562(+0.3\%) & 11.776 & 2.581($-$1.1\%) & 0.1  \\
\hline
\hline
\end{tabular} \\
\end{table}

Figure~\ref{figstrain}(b) shows the three electronic
phases that we can distinguish in bulk Bi under strain: a trivial semiconducting phase, a trivial semimetallic
phase, and a topological semimetallic phase. The overlap between the
electron and hole pockets, or the indirect gap, is caused by the
distortion of the cell with respect to a cubic structure.
It seems, therefore, that bismuth cannot be in a topological
insulating phase because the stress necessary to make it topological
distorts the cell rhombohedrally, distancing it even further
from the cubic symmetry so that Bi becomes more and more metallic.
Figure~\ref{figstrain} demonstrates that, whereas the two
transitions are qualitatively described correctly
by the four approaches, the LDA and GGA predict a large critical strain
that would be very difficult to achieve experimentally.
In particular, the TT transition is predicted to occur at a critical in-plane compressive strain
of 1.9\% and 1.6\%, respectively (see Table~\ref{tabletrans}).
On the contrary, the strain needed to cause the transition is significantly
smaller in $GW$ (0.7\%) and, in particular, in QS$GW$ (0.4\%).
In other words, taking into account structural relaxation,
significantly thicker samples could be grown under
the relatively modest critical strain that is predicted by QS$GW$ 
than under that predicted by DFT.

In conclusion, we propose a QS$GW$ scheme in which the SOC is 
included throughout the whole self-consistent process. 
Our results show that such relativistic QS$GW$ calculations 
are the right \abinitio approach to determine the electronic 
structure and the topological classification of bismuth. They 
predict its two most significant and delicate electronic properties 
in quantitative agreement with experiments:
The direct gap at L, which determines whether Bi is topological or trivial, and the
indirect gap between L and T, which
describes the overlap between the electron and hole pockets and thus
determines the semimetallic character of Bi. 
In response to a recent controversy,\cite{ohtsubo2013} we conclude the following from our
calculations: bulk Bi is a trivial semimetal.

We showed that Bi can undergo electronic phase transitions under rather small lattice strain: 
a trivial SMSC transition under 0.3\% in-plane tensile strain 
and a TT semimetal transition under 0.4\% compressive strain. 
In contrast to LDA and GGA, the critical strains predicted by QS$GW$ are 
attainable by experiments.
The presented results motivate additional experimental efforts to prepare Bi as a topological insulator, by opening band gaps
in slightly strained  Bi. For example, the small strain needed to cause the TT transition concluded by this work,
together with the SMSC transition driven by quantum-size effects in relatively thick films,\cite{xiao2012} 
reveals the potential to observe a topological insulating behavior in bulk-like films of pure Bi.
In fact, first experimental evidences of 3D-like topological thin films of Bi are currently
under discussion.\cite{zhu2014} 

The present results demonstrate that LDA and GGA lack the quantitative
accuracy to predict the critical strain for which the electronic phase
transitions occur. 
We speculate that standard DFT also yields wrong predictions for 
the critical concentration at which Bi experiences a TT transition upon
alloying with other elements (such as in \bisb
alloys\cite{tichovolsky1969,redko1985,teo2008,hsieh2008,hsieh2009}) or for the critical thickness
at which a SMSC transition caused by quantum-size effects takes place. 
Bi is one example, but we believe it is much more general. Other examples 
could include the very large strains necessary to 
induce topological phase transitions in \bise~\cite{young2011,liu2011} and 
in TlBiS$_2$ and TlSbS$_2$.\cite{zhang2015} 
It would thus be desirable to reinvestigate the critical points for these transitions
within the relativistic QS$GW$ method proposed in this work.

\begin{acknowledgments}
We thank Xiaofeng Jin for the very interesting discussions that motivated this work, 
Gregor Mussler for his comments on the experimental growth of Bi, and Gustav Bihlmayer 
for valuable discussions and a critical reading of the manuscript.
This work was supported by the Alexander von Humboldt Foundation through a postdoctoral fellowship, and by the Helmholtz
Association through the Virtual Institute for Topological Insulators (VITI). 
\end{acknowledgments}


\section*{Appendix}


The calculations are carried out with the DFT code {\sc fleur}{}\cite{fleur} and the
$GW$ code {\sc spex}{},\cite{friedrich2010-spex} which are based on 
the all-electron FLAPW formalism. For the valence electrons, space is partitioned into spherical regions around the 
atoms (muffin-tin spheres) and interstitial region between the spheres. 
We use an angular momentum cutoff $l_{\mathrm{max}}=10$ in 
the muffin-tin spheres and a plane-wave cutoff of 4.5~bohr$^{-1}$ in the interstitial region. 
For the DFT calculations, we employ either the Perdew-Zunger\cite{perdew1981} parametrization of the LDA 
exchange-correlation functional or the Perdew-Burke-Ernzerhof\cite{perdew1997} parametrization 
of the GGA. The one-shot $GW$ calculations are always performed using 
the LDA mean-field system as a starting point. 

Due to the very small band gap of bulk Bi (11 to 15~meV for the direct one\cite{maltz1970,vecchi1974,isaacson1969,smith1964} and $-$32 to $-$39 for the indirect one\cite{edelman1975,dinger1973,isaacson1969,smith1964}), all computational parameters entering the $GW$ and 
QS$GW$ calculations are chosen after a systematic study in order to obtain thoroughly converged results. 
An angular momentum cutoff of $l=5$ and a 
linear momentum cutoff of 2.9~bohr$^{-1}$ are employed to construct the mixed product
basis,\cite{kotani2002,friedrich2010-spex} used to represent the dielectric matrix
and the screened interaction. 

The one-shot $GW$ and self-consistent QS$GW$ results for the undistorted
structure are obtained with a 6$\times$6$\times$6 \kv-point mesh sampling the Brillouin zone (Table~\ref{tablegaps} 
and Fig.~\ref{figbands}). 
Fortunately, we found that the QS$GW$ results converge 
faster with respect to the number of \kv points than the results of
one-shot $GW$ so that we could afford to reduce the \kv-point mesh to 
4$\times$4$\times$4 for the QS$GW$ calculations of bismuth under strain. 
For more details about the \kv-point convergence, see below.

To compute the Green function and the polarization function, 500 bands are
used. This corresponds to approximately 130~eV above the Fermi energy. 
The complete fifth shell of Bi is treated as
valence states by the use of local orbitals. In this way, the
contribution of the 5\emph{s}5\emph{p}5\emph{d} states to the Green function and the
electronic screening is fully taken into account. Despite their low
energetic position, inclusion of the 5\emph{s} and 5\emph{p} states effects a change
of 5~meV in the band gap at L.
To describe high-lying states accurately
and to avoid linearization errors,\cite{friedrich2006,friedrich2011,michalicek2013} we complement the basis for the valence electrons for
each atom by  two local orbitals per angular momentum up to $l=3$ with energy parameters far up in the unoccupied states. For the interpolation of the 
band structures in Fig.~\ref{figbands}, maximally localized Wannier functions obtained by the {\sc
wannier90}{} library\cite{wannier90} and a 6$\times$6$\times$6 \kv-point mesh were employed.

We employ the Dirac equation for the core electrons (up to and including the
fourth shell) so that relativistic effects are fully accounted for. 
The core states are thus represented by four-component spinors. This
representation is retained for the evaluation of the core-valence
contribution to the exchange self-energy. (Using averaged core states that are represented in terms of 
non-relativistic $lm$ spherical harmonics instead of the $jm_j$
spinors introduces an
error of 4 and 7~meV in the direct band calculated with one-shot $GW$ and QS$GW$, respectively.)

As a semimetal, bismuth exhibits metallic screening, which is
described technically by the so-called Drude term in the screened interaction $W$.
This term stems from virtual intraband transitions across the Fermi surface.
It can be formulated in a functional form being proportional to the square of the plasma frequency, which 
in turn is evaluated by an integration over the Fermi surface. 
The Drude term can be treated analytically\cite{friedrich2010-spex} and, as long as the
Fermi surface is sufficiently big,
it normally does not pose any numerical problem.
However, bismuth has a very small Fermi surface due to the tiny electron and hole pockets,
which eventually leads to a very sharp ``Drude peak'' in the $GW$ self-energy,
impeding a straight-forward numerical solution of the non-linear quasiparticle equation.
One could also say that, while the Drude term is actually treated correctly, it
gets too much weight because of the finite \kv-point mesh. Therefore, we
neglect the Drude term in our calculations and, instead, simply scale the
head element of $W(\mathbf{k},\omega)$ in the limit
$\mathbf{k}\rightarrow\mathbf{0}$ to enforce metallic screening. This hardly
changes $W$, and we have found that it leads to a very favorable \kv-point
convergence; see below. (In the limit of dense \kv-point sets, the
treatment of the Drude term, which affects only a single \kv point, i.e., the
$\Gamma$ point, becomes immaterial.)

\begin{figure}
\begin{center}
\includegraphics[angle=-90,width=8.2cm]{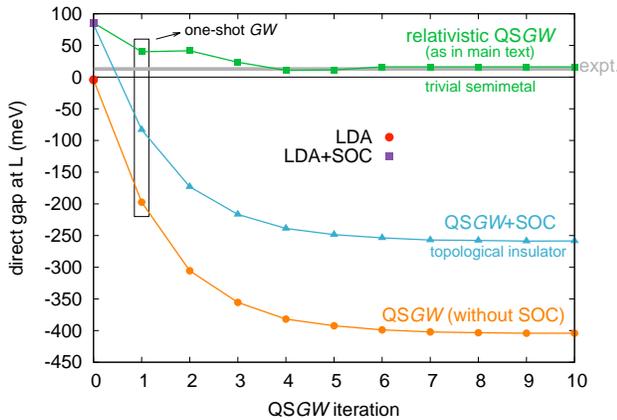}
\caption{\label{conve} (Color online) Convergence of the QS$GW$ direct band
gap at the L point: without SOC (circles), with SOC added \textit{a posteriori} (triangles), and
with the relativistic approach used in the main text (squares). Iteration 0, 1, and 10,
correspond to the LDA, one-shot $GW$, and QS$GW$ values, respectively.}
\end{center}
\end{figure}


In Fig.~\ref{conve}, we show the 
direct band gap at the L point for each iteration in the QS$GW$ self-consistency
procedure,
in particular, we distinguish between calculations
(i) without SOC, (ii) with spin-orbit coupling
added \textit{a posteriori} to a result obtained 
without SOC (QS$GW$+SOC), which is a correction scheme used in most $GW$
calculations so far, 
and (iii) the relativistic approach used in the present work in which spin-orbit coupling is included in the 
calculation of both $G$ and $W$ throughout the whole self-consistent
procedure. 
(In an earlier publication,\cite{aguilera2013-2} 
we have used the notation \fullsoc for the corresponding one-shot 
calculation. By analogy, one could use the notation QS\fullsoc for 
the present self-consistent approach.) 
The \textit{a posteriori} SOC correction leads to unphysical results for Bi, which 
appears as a topological insulator instead of a trivial semimetal. We attribute the 
lack of published $GW$ results of bulk Bi to the crucial and delicate treatment of SOC, 
but also to the difficulties encountered with the semimetallicity of the system exhibiting 
very tiny electron and hole pockets (leading to a very sharp Drude peak in the self-energy, 
as discussed above) and to the smallness of the gaps requiring the calculations to be 
converged to within a few meV.

We have also performed relativistic QS$GW_0$ calculations, 
\ie we keep the $W$ at the LDA level while 
$G$ is updated self-consistently. The converged QS$GW_0$ result 
for the band gap at L differs only by 0.3~meV from the QS$GW$ one. 
This indicates that the screening obtained with LDA is a good approximation 
and that self-consistency induces changes mainly in $G$. Finally, to quantify 
the effects of dynamics, we have also performed self-consistent calculations 
within the static Coulomb-hole screened-exchange (COHSEX)\cite{hedin1965} 
approximation to the $GW$ self-energy (including SOC). In this approximation, 
the changes are more pronounced. With respect to QS$GW$, the
direct gap at L is reduced by 12~meV, nearly making the system 
topological, and the indirect band gap shrinks to -4~meV, nearly making 
the system semiconducting.

\begin{figure*}
\begin{center}
\includegraphics[angle=0,width=16cm]{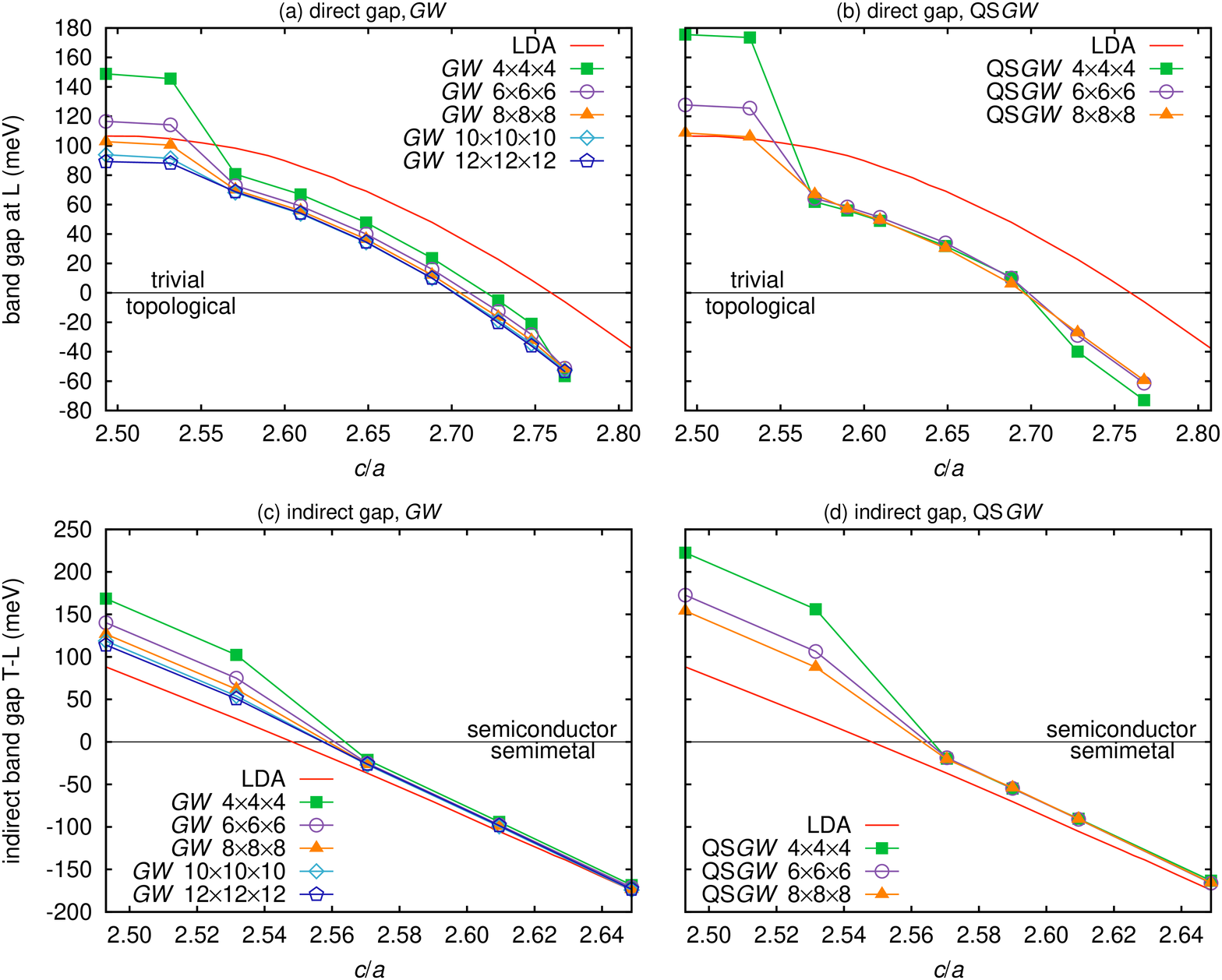}
\caption{\label{kpts} (Color online) LDA, $GW$, and QS$GW$ values of the direct and indirect band 
gaps of bulk Bi as a function of the ratio $c/a$ for different \kv-point sets, but for a set of reduced convergence parameters, see text.}
\end{center}
\end{figure*}

To explain the discontinuities in Figs.~\ref{figstrain}(a) and~\ref{figstrain}(c) and to 
prove their origin as a 
\kv-point convergence issue, we have performed $GW$ and QS$GW$ calculations with increasing number of 
\kv~points 
(Fig.~\ref{kpts}). To allow for these calculations,
we had to reduce some of the parameters in the calculation such as the plane-wave 
cutoff in the interstitial region for the LDA calculations (now 4.0~bohr$^{-1}$), and both the angular momentum 
cutoff ($l=4$) and the linear momentum cutoff (2.6~bohr$^{-1}$) to construct the mixed product basis for the 
$GW$ calculations. We also reduced the number of bands for the $GW$ calculations to 100 
and only 5\emph{d} local orbitals were used in the calculations. 
This obviously reduces the accuracy of the 
band gaps obtained---and therefore they differ from those of the main text--- but it is not our intention 
in this appendix to analyze the quantitative results, only the qualitative behavior due 
to the \kv-point convergence.

From Fig.~\ref{kpts}, we can conclude the following about the \kv-point convergence 
of one-shot and self-consistent $GW$ results: (1) The calculations of the trivial 
semimetallic phases can be considered converged with a 6$\times$6$\times$6 \kv-point 
set for the one-shot $GW$ calculations and 4$\times$4$\times$4 for the
QS$GW$ ones. (2) The semiconducting and topological phases require much
larger \kv-point meshes, but (3) the corresponding results converge
systematically towards an extrapolated 
quadratic (linear) fit to the converged 
trivial semimetallic phases for the direct (indirect) band gap. 
We note that in the case of one-shot $GW$ 
calculations, the discontinuity in the curves due to the \kv-point convergence of the 
topological phases does not appear at the TT transition of $GW$ 
but at that of LDA, because of the use of the LDA mean-field system as
the starting point for the one-shot $GW$ quasiparticle correction.

\bibliography{PRB_bulk-Bi_QSGW-2015-arxiv-v2}
\end{document}